\documentclass[review]{elsarticle}

\usepackage{float}
\usepackage{bm}
\usepackage{amsmath}
\usepackage{amssymb}
\usepackage{graphicx}
\usepackage{setspace}
\usepackage{algorithm}
\usepackage{algpseudocode}
\usepackage{caption}

\usepackage[final]{changes}

\usepackage{lineno,hyperref}
\modulolinenumbers[5]

\journal{Journal of \LaTeX\ Templates}

\begin{document}

\begin{frontmatter}

\title{Folding Simulation of Rigid Origami with Lagrange Multiplier Method}


\author[1]{Yucai Hu}

\author[1,2]{Haiyi Liang\corref{mycorrespondingauthor}}
\cortext[mycorrespondingauthor]{Corresponding author}
\ead{hyliang@ustc.edu.cn}

\address[1]{CAS Key Laboratory of Mechanical Behavior and Design of Materials, Department of Modern Mechanics, University of Science and Technology of China, Hefei, Anhui 230026, China.}
\address[2]{IAT-Chungu Joint Laboratory for Additive Manufacturing, Anhui Chungu 3D printing Institute of Intelligent Equipment and Industrial Technology, Wuhu, Anhui 241200, China.}

\begin{abstract}
Origami crease patterns are folding paths that transform flat sheets into spatial objects. Origami patterns with a single degree of freedom (DOF) have creases that fold simultaneously. More often, several substeps are required to sequentially fold origami of multiple DOFs, and at each substep some creases fold and the rest remain fixed. In this study, we combine the loop closure constraint with Lagrange multiplier method to account for the sequential folding of rigid origami of multiple DOFs, by controlling the rotation of different sets of creases during successive substeps. This strategy is also applicable to model origami-inspired devices, where creases may be equipped with rotational springs and the folding process involves elastic energy. Several examples are presented to verify the proposed algorithms in tracing the sequential folding process as well as searching the equilibrium configurations of origami with rotational springs.
\end{abstract}


\end{frontmatter}


\section{Introduction}
Origami is the art of folding a sheet of paper into artistic three-dimensional object. Origami artists have shown that interesting shape morphing can be attained by folding along crease pattern. Recently, origami goes beyond arts and inspires engineers and scientists to design novel mechanisms, structures, and materials such as deployable space solar sails \cite{miura1985method, zirbel2013accommodating}, medical stents \cite{kuribayashi2006self}, fold cores of sandwich structures \cite{heimbs2013foldcore,ma2018origami}, foldable robots \cite{felton2014method}, meta-materials \cite{silverberg2014using,chen2016topological, lv2014origami}, etc.
To facilitate understanding and design of origami and its inspired derivatives, parametric equations have been derived for regular origami tessellations made up of identical unit cells \cite{wei2013geometric,lv2014origami,yasuda2015reentrant,hanna2014waterbomb}, which would become tedious and even intractable for irregular crease patterns.
The finite element method is applicable in general with the facets modeled by the shell elements and creases by elastic hinges.
It can provide detailed elastic deformation of the facets at the cost of modelling and computational time, which may not be of major concern.
Efficient computational approaches are required to capture the global deformation of the origami.

The bar-and-hinge model can be regarded as a simplified finite element method to account efficiently for folding kinematics when origami is subjected to external forces or torques \cite{schenk2011origami, filipov2017bar, liu2017nonlinear, gillman2018truss, Ghassaei2018FastI}.
In this model, all crease lines are represented by elastic springs, which allow in-plane stretching as well as out-of-plane bending of facets. Schenk and Guest first introduced the bar-and-hinge model for the mechanical analysis of origami where the origami structure is represented by a pin-jointed truss framework, assuming infinitesimal deformation \cite{schenk2011origami}.
Fuchi et al. combined this bar-and-hinge model with topology optimization techniques for the design of origami-based mechanism \cite{Fuchi2015Origami}.
Filipov et al. presented several improvements, including new triangulation schemes for the quadrilateral facets and stiffness parameters obtained from the sheet material, for realistic modeling of origami \cite{filipov2017bar}.
Recently, to solve large deformation problems involving bifurcations and multistability, nonlinear bar-and-hinge models have been developed which synthesize techniques including nonlinear bar elements, rotational springs with penalty near the fully folded state to avoid local penetration, solvers of arch-length method, schemes for tracing the bifurcations, etc \cite{liu2017nonlinear, gillman2018truss}.
Bar-and-hinge model can also be used to simulate the rigid origami folding by projecting the motion into the nullspace of the global constraint matrix, in the limit of infinite large stiffness of bar elements \cite{schenk2011origami,zhang2018folding}.

For origami with stiff facets connected by soft creases, it would be much harder to bend the facets than to fold along the creases.
The rigid origami model is thus proposed in which the facets are assumed rigid and only rotations are allowed along the creases.
The condition of loop closure constraint has to be fulfilled at every interior vertex for all the rigid facets to be compatible with each other during the folding process \cite{belcastro2002modelling,lang2017twists}.
Wu and You have investigated the folding of rigid origami based on the rotating vector model which describes the loop closure constraint using quaternions \cite{wu2010modelling}.
For a given crease pattern, the three-dimensional (3D) folded form of rigid origami can be uniquely determined by fold angles which are the supplementary angles of the dihedral angles between two adjacent facets, see Fig.\ref{fig:vertexfan}.
The kinematic folding of the rigid origami can be described by the parametric equations for origami tessellations made up of identical unit cells or origami with a single vertex.
Wei et al. calculated analytically the Poisson's ratio and stiffness of the Miura-ori tessellation based on the geometry of the unit cell \cite{wei2013geometric}.
Assuming symmetry between the fold angles, Hanna et al. investigated the kinematics of degree-8 Waterbomb base of single vertex with symmetric 8 creases \cite{hanna2014waterbomb}.
Chen et al. presented thorough kinematic folding analysis for degree-6 Waterbomb base of both thin and thick origami \cite{chen2016symmetric}.

Though rigid origami model has been widely accepted in the literature due to its simplicity, most of the previous works are elaborated on the origami tessellations consisting of identical unit cells or the origami base.
To deal with rigid origami of irregular crease pattern, Tachi developed a numerical method based on a modified version of the loop closure constraint developed by belcastro and Hull \cite{belcastro2002modelling}, to simulate the folding motion of rigid origami.
In this algorithm, the infinitesimal rotations of facets were calculated by projecting fold angle changes into the linearized constraint space with the numerical residual compensated by a single Newton-Raphson iteration \cite{tachi2009simulation, tachi2012design}.
However, it may be seen that the folding increment and numerical residual are not strictly controlled for each folding step.

Rigid origami model can be extended to search the elastic equilibrium configurations under the competition between rotational springs mounted at creases with different rest angles, for example, in the case of self-folding mechanisms and robots with smart material actuators mounted at the folds \cite{hawkes2010programmable,ge2014active,rus2018design}.
Brunck et al. derived the covariant energy and the associated geometric constraint (i.e., constant sector angles) for a vertex with $n$ creases using the unit vectors along the creases as the variables \cite{brunck2016elastic}.
Wang and Qiu adopted pseudo-folds to approximate the bent configuration in which the nodal coordinates and the pertinent constraints are expressed with respect to the fold angles \cite{wang2018coupling}.
Both the aforementioned two studies focus on the rigid origami of single vertex.
The ground structure, a potential structure equipped with a sufficient number of pseudo-folds, has been employed for the origami design by considering the stiffness of the crease rotational springs as the weight function \cite{fuchi2013origami}.

Most of existing studies focus on origami of single DOF, such as Miura-ori that folds simultaneously.
In many cases, crease pattern is of multiple DOFs and usually requires sequential folding, so that it takes several distinct steps to fold an origami and at each substage some creases fold and the rest remain fixed.
This complicates the folding analysis, and leaves analytical solution out of reach.
In this paper, we pay attention to sequential folding analysis of rigid origami
when the folding sequence is known.
The proposed algorithm is presented for the folding analysis of the rigid origami without inner holes.
In the following sections, the rigid origami model with the fold angles as the variables is reviewed first.
Then, the Lagrange multiplier method is applied to control a subset of creases which drives the folding of the rigid origami.
The numerical residual of each folding step is eliminated by the Newton-Raphson method, and thus the fold angles yield valid configuration.
In addition, by considering the rotational springs at the creases, an algorithm based on the Lagrange multiplier method is presented to search the equilibrium configuration of the rigid origami.
It is shown that the projection method by Tachi \cite{tachi2009simulation} agrees with a special case of the current physical model. The algorithms are then verified by several examples.

\section{Rigid origami model}

\begin{figure}
	\begin{centering}
		\includegraphics[width=12cm]{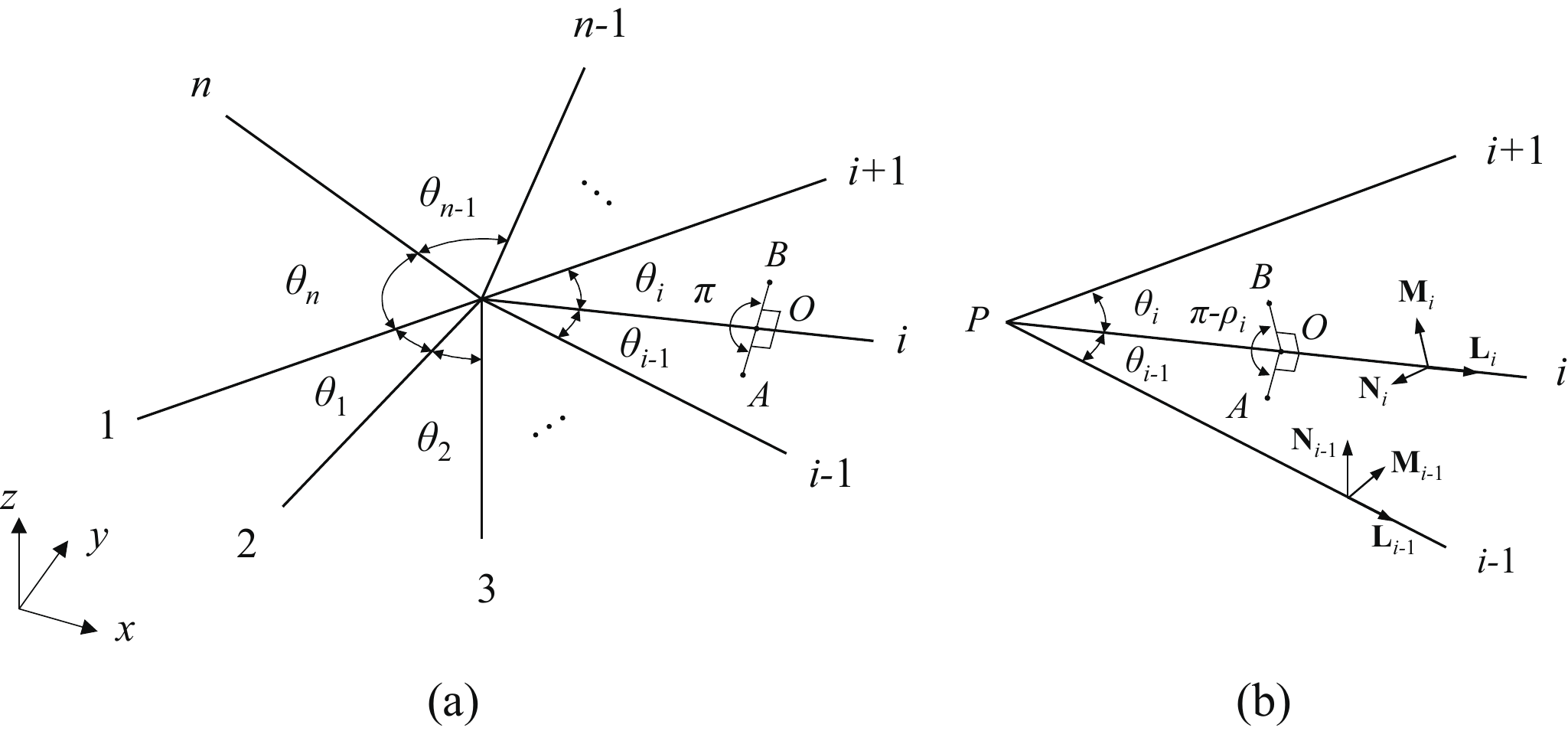}
		\par\end{centering}
	\caption{\label{fig:vertexfan}(a) A vertex with $n$ creases before folding.
		$\theta_{1}+\theta_{2}+\cdots+\theta_{n}=2\pi$. AOB is a straight
		line perpendicular to crease $i$ with both A and B on the origami
		and $\angle AOB=\pi$. (b) AOB is kinked after folding. $\angle AOB$ becomes
		$\pi-\rho_{i}$, where $\rho_{i}$ is the fold angle of crease $i$.
		The triplet, $\textbf{L}_{i}$, $\textbf{M}_{i}$ and $\textbf{N}_{i}$,
		defines a local Cartesian coordinate system for the sector facet $(i,i+1)$:
		$\textbf{L}_{i}$ is the unit vector along crease $i$; $\textbf{N}_{i}$
		is the unit normal of the sector facet $(i,i+1)$ and $\textbf{M}_{i}=\textbf{N}_{i}\times\textbf{L}_{i}$
		where ``$\times$'' is the cross product. The indexes ($i$-1) and
		($i$+1) are to be interpreted cyclically.}
\end{figure}

The rigid origami model of Tachi \cite{tachi2009simulation, tachi2012design}
is reviewed in this section for completeness. The origami is commonly
designated by the crease pattern consisting of vertexes and creases.
Vertexes are points on the origami paper and each crease is a line
joining two neighboring vertexes along which origami paper is folded.
Fig.\ref{fig:vertexfan}(a) shows an isometric view of a typical vertex $P$ with
$n$ creases joining it on an unfolded or flat origami paper. The
creases are numbered from 1 to $n$ anticlockwise. Creases $i$ and
$i+1$ define the sector angle $\theta_{i}$. It is trivial that the
sum of the $n$ sector angles is $2\pi$. Before folding, AOB is a
straight line perpendicular to crease $i$ on the origami paper, i.e.,
$\angle AOB$ is $\pi$ as shown in the figure. After folding, $\angle AOB$
would be denoted as $\pi-\rho_{i}$ in which $\rho_{i}$ is the fold
angle of crease $i$. For the rigid facets to be compatible with each
other, the loop closure constraint around the vertex $P$ is

\begin{equation}
\textbf{F}(\bm{\rho}_{P})=\bm{\chi}_{1,2}\bm{\chi}_{2,3}\cdots\bm{\chi}_{n,1}=\textbf{I}_{3}\ .\label{eq:LoopCons}
\end{equation}
where
\begin{equation}
\bm{\chi}_{i-1,i}=\begin{bmatrix}\cos\theta_{i-1} & -\sin\theta_{i-1} & 0\\
\sin\theta_{i-1} & \cos\theta_{i-1} & 0\\
0 & 0 & 1
\end{bmatrix}\begin{bmatrix}1 & 0 & 0\\
0 & \cos\rho_{i} & -\sin\rho_{i}\\
0 & \sin\rho_{i} & \cos\rho_{i}
\end{bmatrix}\ ;\label{eq:single_rot}
\end{equation}
$\textbf{I}_{m}$ is the $m\times m$ identity matrix and $\bm{\rho}_{P}=\{\rho_{1},\rho_{2},\cdots,\rho_{n}\}$
is the vector containing the $n$ fold angles. The derivation of Eq.(\ref{eq:LoopCons})
is given in Appendix A. As the facets bounded by adjacent creases
are rigid, the sector angles are constants and Eq.(\ref{eq:LoopCons})
are nonlinear constraints on the fold angles. It can be shown that
the matrix $\frac{\partial\textbf{F}}{\partial\rho_{i}}$ is antisymmetric
for compatible fold angles (see Appendix A for the proof), i.e.,
\begin{equation}
\left.\frac{\partial\textbf{F}}{\partial\rho_{i}}\right|_{\bm{\rho}_{P}}=\begin{bmatrix}0 & -c_{i} & b_{i}\\
c_{i} & 0 & -a_{i}\\
-b_{i} & a_{i} & 0
\end{bmatrix}\quad\text{for}\quad i=1,2,\cdots,n\ .\label{eq:asymF}
\end{equation}

For each folding step in simulating the folding process, the iteration
starts with an approximate solution $\bm{\rho}_{P}^{i}=\{\rho_{1}^{i},\rho_{2}^{i},\cdots,\rho_{n}^{i}\}$
and the associated residual is
\begin{equation}
\textbf{R}(\bm{\rho}_{P}^{i})=\textbf{F}(\bm{\rho}_{P}^{i})-\textbf{I}_{3}\ .\label{eq:residual}
\end{equation}
The aim of iteration is to find an increment or infinitesimal folding,
$\Delta\bm{\rho}_{P}$, such that the residual for $\bm{\rho}_{P}^{i+1}=\text{\ensuremath{\bm{\rho}}}_{P}^{i}+\Delta\bm{\rho}_{P}$
is eliminated. Substitute $\bm{\rho}_{P}^{i+1}$ into Eq.(\ref{eq:residual})
and expand the residual about $\bm{\rho}_{P}^{i}$ as,

\begin{equation}
\textbf{R}(\bm{\rho}_{P}^{i+1})=\textbf{F}(\bm{\rho}_{P}^{i}+\Delta\bm{\rho}_{P})-\textbf{I}_{3}\simeq\textbf{F}(\bm{\rho}_{P}^{i})+\sum_{j=1}^{n}\left.\frac{\partial\textbf{F}}{\partial\rho_{j}}\right|_{\bm{\rho}_{P}^{i}}\Delta\rho_{j}-\textbf{I}_{3}\ .\label{eq:residual_i1}
\end{equation}
Thus,
\begin{equation}
\sum_{j=1}^{n}\left.\frac{\partial\textbf{F}}{\partial\rho_{j}}\right|_{\bm{\rho}_{P}^{i}}\Delta\rho_{j}=-\{\textbf{F}(\bm{\rho}_{P}^{i})-\textbf{I}_{3}\}=-\textbf{R}(\bm{\rho}_{P}^{i})\ .\label{eq:frq}
\end{equation}
It is assumed that the increment is infinitesimal and $\bm{\rho}_{P}^{i}$
is infinitesimally perturbed from the valid state, thus the antisymmetry
of the derivative matrix is approximately retained. Thus, only three
of the nine equations in Eq.(\ref{eq:frq}) are independent. Let $a_{j}$,
$b_{j}$ and $c_{j}$ be, respectively, the entries with the indexes
(3,2), (1,3) and (2,1) in $\left.\frac{\partial\textbf{F}}{\partial\rho_{j}}\right|_{\bm{\rho}_{P}^{i}}$,
the independent equations of (\ref{eq:frq}) can be written as
\begin{equation}
\begin{bmatrix}a_{1} & a_{2} & \cdots & a_{n}\\
b_{1} & b_{2} & \cdots & b_{n}\\
c_{1} & c_{2} & \cdots & c_{n}
\end{bmatrix}\begin{bmatrix}\Delta\rho_{1}\\
\Delta\rho_{2}\\
\vdots\\
\Delta\rho_{n}
\end{bmatrix}=-\begin{bmatrix}\textbf{F}_{3,2}\\
\textbf{F}_{1,3}\\
\textbf{F}_{2,1}
\end{bmatrix}\label{eq:slc}
\end{equation}
where $\textbf{F}_{i,j}$ is the $(i,j)$-entry of the matrix $\textbf{F}$.

For general crease pattern consisting of $N_{Vi}$ interior vertexes,
$N_{Ei}$ creases (interior edges) and no holes, Eq.(\ref{eq:slc})
can be formed for every interior vertex with the incident fold angles.
Thus, there are $3N_{Vi}$ constraints on the $N_{Ei}$ fold angles.
Let $\bm{\rho}^{i}$ be the vector of approximate fold angles renumbered
globally, Eq.(\ref{eq:slc}) for every interior vertex can be collected
and expressed as

\begin{equation}
\textbf{C}\Delta\bm{\rho}=-\textbf{r}\label{eq:glcR}
\end{equation}
where $\textbf{C}$ is the global linearized constraint matrix of
dimension $3N_{Vi}\times N_{Ei}$; $\Delta\bm{\rho}$ is the vector
of fold angle increment and $\textbf{r}$ is the vector of residuals.
Note that both $\textbf{C}$ and $\textbf{r}$ are evaluated at $\bm{\rho}^{i}$.
If $\bm{\rho}^{i}$ is a vector of valid fold angles, $\textbf{r}$
is zero and Eq.(\ref{eq:glcR}) reduces to the linearized constraint
on the infinitesimal fold angles. Starting with a valid state and
the vector of intended fold angle changes $\Delta\bm{\rho}_{0}$,
Tachi introduced the Euler method to simulate the folding process
by projecting $\Delta\bm{\rho}_{0}$ into the nullspace of $\textbf{C}$
and compensating the accumulated numerical error \cite{tachi2009simulation}:

\begin{equation}
\Delta\bm{\rho}=[\textbf{I}_{N_{Ei}}-\textbf{C}^{+}\textbf{C}]\Delta\bm{\rho}_{0}-\textbf{C}^{+}\textbf{r}\label{eq:TachiPJ}
\end{equation}
where $\textbf{C}^{+}$ denotes the Moore-Penrose pseudoinverse of
$\textbf{C}$. However, it is hard to control the folding increment
using the above projection method.

\section{Sequential folding using Lagrange multiplier method\label{sec:control_crease}}

Due to the loop closure constraint around every inner vertex, the
folding motion of the origami can usually be driven by folding a subset
of creases. For instance, by controlling anyone of the fold angles
of the Miura-ori fold, the entire origami can be folded simultaneously
since the structure is a mechanism with a single DOF. For general
origami, the DOF of the origami in the partially folded state can
be obtained from the nullity of the linearized constraint matrix $\textbf{C}$.
Let $I_{c}$ be the set of controlled fold angles whose values are
prescribed, i.e., $\rho_{j}=\bar{\rho}_{j}$ for $j\in I_{c}$. The
folding process consists of a number of folding steps. For the $i$-th
folding step, the increment for the controlled angle $\rho_{j}$ is
$\Delta\rho_{j}=\bar{\rho}_{j}^{i+1}-\bar{\rho}_{j}^{i}=f_{j}$ where
$f_{j}$ is self-defined. Except the controlled fold angles, the other
components in $\bm{\rho}^{i}$ are infinitesimally perturbed from
the valid state. A direct approach is to eliminate the controlled
fold angles from Eq.(\ref{eq:glcR}) which, however, alters the structure
of the system matrix and can be inconvenient when different sets of
creases are controlled in multiple folding stages. In the following,
the Lagrange multipliers $\bm{\lambda}=\{\lambda_{j}\}$ are introduced
for the controlled fold angles and the functional is
\begin{equation}
\varPi(\Delta\bm{\rho},\bm{\lambda})=\dfrac{1}{2}(\textbf{C}\Delta\bm{\rho}+\textbf{r})^{\text{T}}(\textbf{C}\Delta\bm{\rho}+\textbf{r})+\sum_{j\in I_{c}}\lambda_{j}(\Delta\rho_{j}-f_{j}) \ .
\end{equation}
The solution is given by the point where the derivatives with respect
to $\Delta\bm{\rho}$ and $\bm{\lambda}$ are zero:
\begin{equation}
\begin{bmatrix}\textbf{C}^{\text{T}}\textbf{C} & \textbf{A}\\
\textbf{A}^{\text{T}} & \textbf{0}
\end{bmatrix}\begin{Bmatrix}\Delta\bm{\rho}\\
\bm{\lambda}
\end{Bmatrix}=\begin{Bmatrix}-\textbf{C}^{\text{T}}\textbf{r}\\
\textbf{f}
\end{Bmatrix}\label{eq:Lg_fm}
\end{equation}
in which $\textbf{A}=[\textbf{e}_{i}]_{i\in I_{c}}$ with $\textbf{e}_{i}$
being a column vector of length $N_{Ei}$ with 1 in the $i$-th position
and 0 in every other position and $\textbf{f}=\{f_{j}\}$. As the
system can be under-determined, the minimum length solution for Eq.(\ref{eq:Lg_fm})
is
\begin{equation}
\begin{Bmatrix}\Delta\bm{\rho}\\
\bm{\lambda}
\end{Bmatrix}=\begin{bmatrix}\textbf{C}^{\text{T}}\textbf{C} & \textbf{A}\\
\textbf{A}^{\text{T}} & \textbf{0}
\end{bmatrix}^{+}\begin{Bmatrix}-\textbf{C}^{\text{T}}\textbf{r}\\
\textbf{f}
\end{Bmatrix}\ .\label{eq:Lg_fmS}
\end{equation}
It should be remarked that the vector of updated fold angles $\bm{\rho}^{i}+\Delta\bm{\rho}$
does not satisfy the nonlinear loop closure constraint in Eq.(\ref{eq:LoopCons})
exactly in general. By updating $\textbf{C}$ and $\textbf{r}$, and
replacing $\textbf{f}$ by $\textbf{0}$ so as to keep the controlled
fold angles unchanged, Eq.(\ref{eq:Lg_fmS}) is looped to reduce the
residual $\textbf{r}$ until that the convergence tolerance is satisfied
for the $i$-th folding step. The procedure for a single folding step
is summarized in Algorithm \ref{alg:one_folding} which can be repeated
to obtain the whole folding motion.

For each folding step, the resultant fold angles are compatible and
the 3D folded form of the origami can be visualized by calculating
the coordinates of the vertexes based on the crease pattern and fold
angles. The procedure for calculating the 3D folded form is presented
in the Appendix B.
Many origami artworks involve a sequence of folding
stages or sequential folding.
With all the creases and the folding sequence specified \emph{a priori},
the simulation can be processed by updating $I_{c}$ and
$\textbf{f}$ for each stage.
More details are exposed in the example \ref{subsec:crane} by considering the sequential folding of a crane.

\begin{algorithm}
\caption{\label{alg:one_folding}A single folding step driven by controlled
creases}

\begin{algorithmic}[1]   		
\Require{Sector angles at all vertexes, $\bm{\theta}$; Current fold angles, $\bm{\rho}^{i}$; The specified increment for ${\Delta\rho_j}$ with ${j\in I_c}$, $\textbf{f}$; Tolerance for the residual error, $\epsilon=10^{-9}$.}
\Ensure{The vector of fold angles for the next step, $\bm{\rho}^{i+1}$.}
\State {Calculate $\textbf{C}$ and $\textbf{r}$ at $\bm{\rho}^{i}$;}
\State {Calculate the increment $\Delta\bm{\rho}$ from (\ref{eq:Lg_fm});}
\State {Update $\bm\rho^{i+1} = \bm\rho^{i}+\Delta\bm{\rho}$ and $\textbf{r}$;}
\While {$(\| \textbf{r} \| / (3N_{Vi}) \geq \epsilon )$}   		
\State Update $\textbf{C}$ at $\bm\rho^{i+1}$;   		
\State Solve $ \Delta\bm{\rho}$ from (\ref{eq:Lg_fm}) with $\textbf{f}=\textbf{0}$ ;   		
\State $\bm\rho^{i+1} \gets \bm\rho^{i+1}+\Delta\bm{\rho}$; update $\textbf{r}$ at $\bm\rho^{i+1}$;		
\EndWhile 		
\end{algorithmic}  		
\end{algorithm}

\section{Elastic folding using Lagrange multiplier method\label{sec:folding_method}}

Rigid origami can be elastically folded and multistable by mounting rotational
springs at crease lines. In fact, various actuators such as shape memory alloy and shape memory polymer \cite{hawkes2010programmable,rus2018design} have been employed to actuate the folding of origami inspired structures. Considering rotational springs at the creases,
the elastic energy of the origami structure is

\begin{equation}
U(\bm{\rho})=\dfrac{1}{2}\sum_{i=1}^{N_{Ei}}k_{i}(\rho_{i}-\tilde{\rho}_{i})^{2}\label{eq:objective_U}
\end{equation}
where $k_{i}$ and $\tilde{\rho}_{i}$ are, respectively, the rotational
spring stiffness and rest angle of crease-$i$ and the former is $k_{i}=kL_{i}$
in which $L_{i}$ is the length of the crease and the constant $k$
is the stiffness per unit length. When the fold angle $\bm{\rho}$ is
increased by $\Delta\bm{\rho}$, the energy increment is
\begin{equation}
\Delta U=U(\bm{\rho}+\Delta\bm{\rho})-U(\bm{\rho})=\dfrac{1}{2}\Delta\bm{\rho}^{\text{T}}\textbf{H}\Delta\bm{\rho}+\textbf{d}^{\text{T}}\Delta\bm{\rho}\label{eq:U_increm}
\end{equation}
where $\textbf{H}=diag.\{k_{1},k_{2},\cdots,k_{N_{Ei}}\}$ and
\begin{equation}
\textbf{d}=[k_{1}(\rho_{1}-\tilde{\rho}_{1}),k_{2}(\rho_{2}-\tilde{\rho}_{2}),\cdots,k_{N_{Ei}}(\rho_{N_{Ei}}-\tilde{\rho}_{N_{Ei}})]^{\text{T}}=\dfrac{\partial U}{\partial\bm{\rho}}\label{eq:dUdrho}
\end{equation}
which is essentially a vector of internal moments along the creases.
The aim is to find $\Delta\bm{\rho}$ such that $\Delta U$ is minimized
while subjected to the loop closure constraint of the rigid origami,
see Eq.(\ref{eq:LoopCons}). By considering the linearized constraint
on the increment in Eq.(\ref{eq:glcR}), the minimization of $\Delta U$
subjected to nonlinear constraints is transformed into a classical
quadratic problem \cite{nocedal2006optimization}. Introducing the
loop closure constraints on internal vertices by the Lagrange multipliers
$\bm{\lambda}$, the functional on the infinitesimal increment can
be written as

\begin{equation}
\Delta\varPi_{U}=\dfrac{1}{2}\Delta\bm{\rho}^{\text{T}}\textbf{H}\Delta\bm{\rho}+\textbf{d}^{\text{T}}\Delta\bm{\rho}+\bm{\lambda}^{\text{T}}(\textbf{C}\Delta\bm{\rho}+\textbf{r})\ .\label{eq:opti_func}
\end{equation}
Variations of (\ref{eq:opti_func}) with respect to $\Delta\bm{\rho}$
and $\bm{\lambda}$ yields

\begin{equation}
\begin{bmatrix}\textbf{H} & \textbf{C}^{\text{T}}\\
\textbf{C} & \textbf{0}
\end{bmatrix}\begin{Bmatrix}\Delta\bm{\rho}\\
\bm{\lambda}
\end{Bmatrix}=-\begin{Bmatrix}\textbf{d}\\
\textbf{r}
\end{Bmatrix}\ .\label{eq:Lg_eng}
\end{equation}
The minimum length solution for Eq.(\ref{eq:Lg_eng}) is

\begin{equation}
\begin{Bmatrix}\Delta\bm{\rho}\\
\bm{\lambda}
\end{Bmatrix}=-\begin{bmatrix}\textbf{H} & \textbf{C}^{\text{T}}\\
\textbf{C} & \textbf{0}
\end{bmatrix}^{+}\begin{Bmatrix}\textbf{d}\\
\textbf{r}
\end{Bmatrix}\ .\label{eq:pinv_eng}
\end{equation}
In the case of rigid origami with triangular facets with no holes,
it can be shown that $3N_{Vi}\leq N_{Ei}$ \cite{tachi2010geometric}.
Assuming that all the linearized constraints in Eq.(\ref{eq:glcR})
are independent, i.e., $\text{Rank}(\textbf{C})=3N_{Vi}$, explicit
expression for the inverse matrix in Eq.(\ref{eq:pinv_eng}) can be
derived and the increment is (see Section 16.2 of the book \cite{nocedal2006optimization})

\begin{equation}
\Delta\bm{\rho}=-(\textbf{H}^{-1}-\textbf{G}\textbf{C}\textbf{H}^{-1})\textbf{d}-\textbf{G}\textbf{r}\quad\text{with}\quad\textbf{G=}\textbf{H}^{-1}\textbf{C}^{\text{T}}(\textbf{C}\textbf{H}^{-1}\textbf{C}^{\text{T}})^{-1}\ .\label{eq:increm_eng}
\end{equation}
In the special case that all springs are of the same stiffness, i.e.,
$k_{i}=k_{0}$ for $i=1,2,\cdots N_{Ei}$, the diagonal matrix $\textbf{H}$
reduces to the identity matrix $\textbf{I}_{N_{Ei}}$ multiplied by
$k_{0}$. Thus, we have $\textbf{G}=\textbf{C}^{\text{T}}(\textbf{C}\textbf{C}^{\text{T}})^{-1}=\textbf{C}^{+}$
and Eq.(\ref{eq:increm_eng}) can be simplified as

\begin{equation}
\Delta\bm{\rho}=-\frac{1}{k_{0}}(\textbf{I}_{N_{Ei}}-\textbf{C}^{+}\textbf{C})\textbf{d}-\textbf{C}^{+}\textbf{r}\ .\label{eq:increm_pj}
\end{equation}
As $\textbf{d}$ is the gradient of the energy, Eq.(\ref{eq:increm_pj})
means that the vector of increment is given by the fastest decreasing
direction of the system energy projected to the linearized constraint
space. It can be seen that Eq.(\ref{eq:increm_pj}) agrees with the
projection method in Eq.(\ref{eq:TachiPJ}) (also see Eq.(14) of the
reference \cite{tachi2009simulation}) when the internal moment $\textbf{d}$
is treated as the vector of intended fold angle changes. Here, the
physical meaning is more clear and Eq.(\ref{eq:pinv_eng}) should
be used for general cases.

It should be remarked that $\Delta\bm{\rho}$ suggests the optimal
direction to decrease the system energy while satisfying the linearized
closure constraint. The step length of the searching should, however,
be restricted such that the changes in the fold angles are still infinitesimal.
In this light, a step length factor $c$ is introduced and

\begin{equation}
\bm{\rho}^{i+1}=\bm{\rho}^{i}+c\frac{\Delta\bm{\rho}}{\max(\text{abs}(\Delta\bm{\rho}))}\label{eq:step_increm_eng}
\end{equation}
where $\max(\text{abs}(\Delta\bm{\rho}))$ extracts the largest magnitude
of the components in the fold angle increment vector. The factor enforce
that the largest change in a fold angle is $c$ during a single folding
step and we restrict that $c\leq\pi/36$. Similar to the discussion
following Eq.(\ref{eq:Lg_fmS}) in Subsection \ref{sec:control_crease},
the fold angle $\bm{\rho}^{i+1}$ does not exactly fulfill the nonlinear
loop consistency constraint in Eq.(\ref{eq:LoopCons}). The residual
$\textbf{r}$ is to be compensated with iterations of $\Delta\bm{\rho}=-\textbf{C}^{+}\textbf{r}$.
Since the length of the fold step is kept small, the energy will typically
decrease monotonously in the initial folding steps. A fold angle which
increases/decreases monotonously before reaching the local energy
minimum is chosen as the characteristic fold angle and indicated by
$\rho_{a}$. When the increment of the characteristic angle is reversed,
i.e., $\Delta\rho_{a}(\rho_{a}^{i}-\rho_{a}^{i-1})<0$, the local
minimum state is within the region bounded by $\rho_{a}^{i}$ and
$\rho_{a}^{i}+c\Delta\rho_{a}/\max(\text{abs}(\Delta\bm{\rho}))$.
To converge to the local minimum, the step length factor $c$ is divided
by two, i.e., $c\leftarrow c/2$, and the algorithm is summarized
in Algorithm \ref{alg:one_folding_eng}.

\begin{algorithm}
\caption{\label{alg:one_folding_eng}Folding driven by the rotational springs}

\begin{algorithmic}[1]   		
\Require{Sector angles at all vertexes, $\bm{\theta}$; Initial fold angles, $\bm{\rho}^{0}$; Spring stiffness for the crease-$i$, $k_i$; Characteristic fold angle, $\rho_a$; Initial step length factor, $c_0$; Tolerances, $\epsilon_1$ and $\epsilon_2$.}
\Ensure{Folding states $\bm{\rho}^{i}$ and the converged fold angles for the local minimum.}
\State {$\ i \gets 0, \ c \gets c_0, \ \bm{\rho}^{1} \gets \bm{\rho}^{0}$};
\While {$c>\epsilon_1$ and $i<$ specified maximum increment number}
\State {$i \gets i+1$};
\State {Calculate $\textbf{C}$ and $\textbf{r}$ at $\bm{\rho}^{i}$};
\State {Calculate the increment $\Delta\bm{\rho}$ from (\ref{eq:pinv_eng})};
\If{$i>2$ and $\Delta{\rho_a}(\rho_{a}^{i}-\rho_{a}^{i-1})<0$}
\State $c \gets c/2$;
\EndIf
\State {$\bm\rho^{i+1} \gets \bm\rho^{i}+c\Delta\bm{\rho}/\max(\text{abs}(\Delta\bm{\rho}))$ and update $\textbf{r}$};
\While {$(\| \textbf{r} \| / (3N_{Vi}) \geq \epsilon_2 )$}   		
\State {Update $\textbf{C}$ at $\bm{\rho}^{i+1}$};   		
\State {$ \Delta\bm{\rho} \gets -\textbf{C}^{+}\textbf{r}$; \  $\bm\rho^{i+1} \gets \bm\rho^{i+1}+\Delta\bm{\rho}$};
\State {Update $\textbf{r}$ at $\bm\rho^{i+1}$};		
\EndWhile	
\EndWhile 	
\end{algorithmic}  		
\end{algorithm}

\section{Simulations of sequential/elastic folding}

In this section, the algorithms proposed in the previous sections \ref{sec:control_crease} and \ref{sec:folding_method} are validated by four examples of controlled origami folding, where the advantages of combining the loop closure constrains and Lagrange multiplier method are demonstrated.
The first two examples are dedicated to the sequential folding algorithm, i.e., Algorithm \ref{alg:one_folding}.
In subsection \ref{subsec:Miura}, as an example of single DOF, the simultaneous folding of Miura-ori is achieved by solely controlling one crease in one simulation step, and the folding kinetics is compared with the analytic solutions. Subsection \ref{subsec:crane} illustrates the sequential folding of origami crane, an typical origami example of multiple DOFs, using three folding substeps, where different sets of creases are successively controlled with Lagrange multiplier. The last two examples in subsections \ref{subsec:Waterbomb_base} and \ref{subsec:WBT}, folding simulations of Waterbomb and Waterbomb tessellation, show the elastic bistability or equilibrium configuration can be reached iteratively using Algorithm \ref{alg:one_folding_eng}, when rigid origami are equipped with rotational springs at creases.

\subsection{Folding of Miura-ori \label{subsec:Miura}}

\begin{figure}
	\begin{centering}
		\includegraphics[width=12cm]{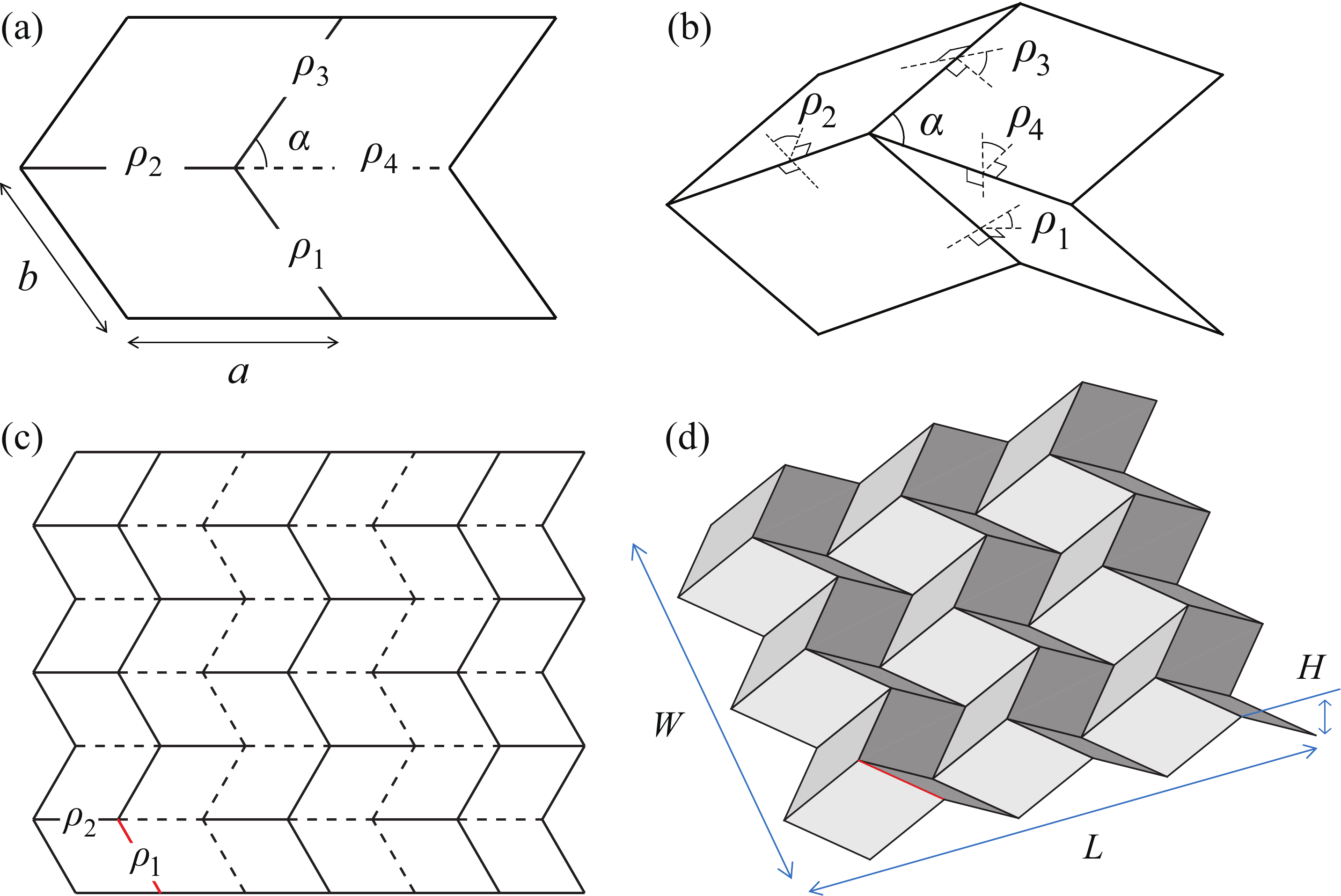}
		\par\end{centering}
	\caption{\label{fig:miura}(a) Parameters for the Miura-ori unit cell; mountain
		and valley creases are indicated by the
		black solid and black dashed
		lines, respectively;
		(b) partially folded configuration of the unit cell,
			the fold angle of valley crease is positive, i.e, $\rho_4>0$,
			whilst those of mountain creases are negative;
		(c) crease pattern for the Miura-ori fold consisting
		of 3$\times$3 unit cells with the parameters $a=b=1$ and $\alpha=\pi/3$;
		(d) the dimensions of the folded form.
	In (c) and (d), the fold angle of the red crease is controlled with the Lagrange multiplier.}
\end{figure}

The Miura-ori unit is made up of four identical parallelograms
characterized by the parameters $a$, $b$ and $\alpha$, see Fig.\ref{fig:miura}(a) and (b).
The folding motion of a Miura-ori fold with 3$\times$3 unit cells
is simulated by controlling the angle $\rho_{1}$ as shown in Fig.\ref{fig:miura}(c).
For the planar state with all fold angles equal 0, the third equation
in Eq.(\ref{eq:slc}) will degenerate as $c_{i}$, the (2,1)-entry
of the derivative matrix in Eq.(\ref{eq:asymF}), equals 0 \cite{tachi2012design}.
This reflects the fact that several folded forms are permissible as
the mountain and valley crease assignment is missing. In the simulation,
the initial fold angles are prescribed with small values, for instance
$\pm1^{\circ}$, whilst the positive and negative signs are assigned
for the valley and mountain fold angles, respectively. The numerical
error can be eliminated by the iterations in the Algorithm \ref{alg:one_folding}.

\begin{figure}
	\begin{centering}
		\includegraphics[width=12cm]{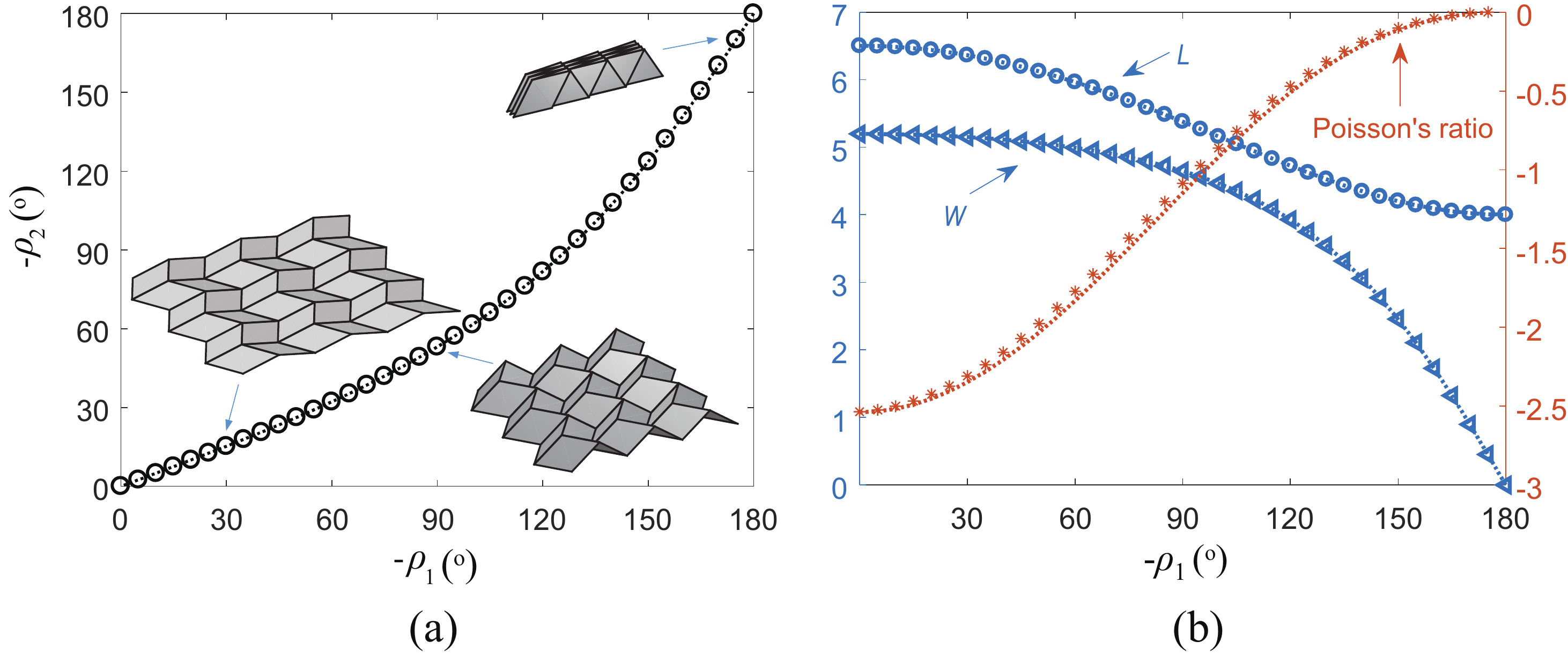}
		\par\end{centering}
	\caption{\label{fig:miura_folding}(a) The fold angle $\rho_{2}$ versus $\rho_{1}$;
		the insets are frames for$\rho_{1}=-30^{\circ},-90^{\circ}$ and $-175^{\circ}$;
		(b) the length $L$, width $W$ and Poisson's ratio versus $\rho_{1}$.
		For each variable, the markers are the results obtained form the simulation
		whilst the dash lines are the analytical solutions\cite{lv2014origami}.}
\end{figure}

During the simulation, the controlled angle $\rho_{1}$
, see the red crease in Fig.\ref{fig:miura}(c) and (d),
decreased from $0^{\circ}$ to $-180^{\circ}$ by $-5^{\circ}$ in every folding
step. Fig.\ref{fig:miura_folding}(a) shows the $\rho_{2}$ versus
$\rho_{1}$ which is in agreement with the analytical solution (the
dash line): $\tan(\rho_{2}/2)=\cos(\alpha)\tan(\rho_{1}/2)$ \cite{evans2015rigidly}.
The insets in Fig.\ref{fig:miura_folding}(a) are the folded form
with $\rho_{1}=-30^{\circ},-90^{\circ}$ and $-175^{\circ}$. The
dimensions of the folded form, see Fig.\ref{fig:miura}(d), are measured
at the end of each folding step. Fig.\ref{fig:miura_folding}(b) shows
the change of width and length of the 3$\times$3 Miura-ori fold during
the folding process. In fact, the length and width agree exactly with
the pertinent analytical predictions as the numerical error is negligible
at the end of each folding step. The in-plane Poisson's ratio, defined
as $\nu_{LW}=-\big(\frac{dL}{L}\big)/\big(\frac{dW}{W}\big)$ \cite{wei2013geometric},
is calculated numerically as $\nu_{LW}^{h}=-\big(\frac{L_{i+1}-L_{i}}{L_{i}}\big)/\big(\frac{W_{i+1}-W_{i}}{W_{i}}\big)$
where the subscript $i$ and $i+1$ correspond to the number of folding
step. It can be seen that the numerical predictions for Poisson's
ratio agree with the analytical solution and the accuracy can be improved
if a smaller folding step is used. This example validates the effectiveness
of Algorithm \ref{alg:one_folding} which is of potential to trace
exactly the geometric change of general rigid origami. It is worth
mentioning that the fold angle $\rho_{1}$ is folded exactly to the
specified angles during the folding process which presents challenge
for the projection method in the reference \cite{tachi2009simulation}.

\subsection{Sequential folding of origami crane\label{subsec:crane}}

\begin{figure}
	\begin{centering}
		\includegraphics[width=10cm]{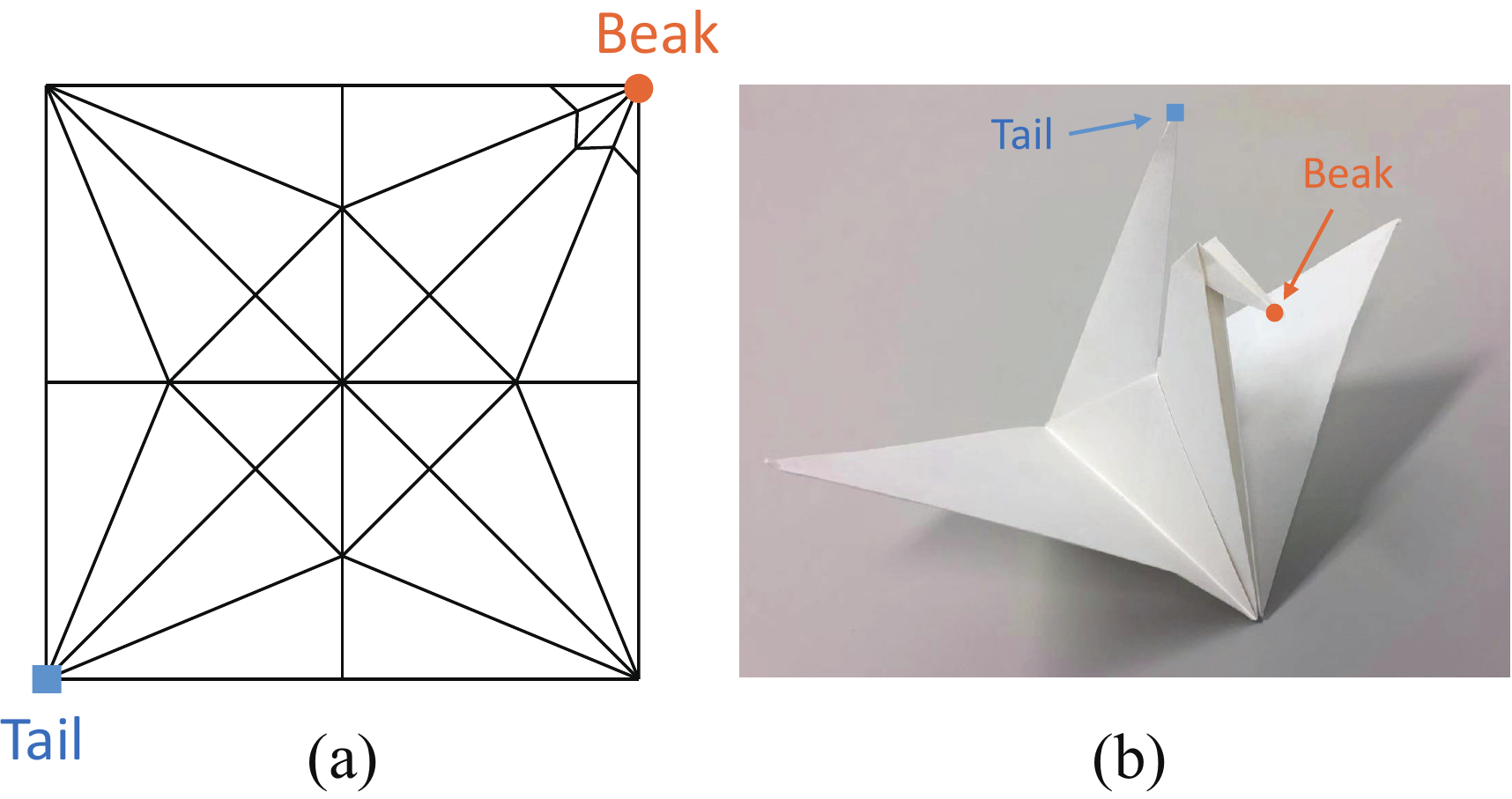}
		\par\end{centering}
	\caption{\label{fig:craneCP}(a) Creases and boundary edges for the crane; (b) Photo of the folded paper crane. The beak and tail of the crane are indicated by the circle and square, respectively.}
\end{figure}

When folding the classical origami crane from a blank sheet by hand, a sequence of folding steps are needed to attain the final shape due to its multiple DOFs of the crease pattern. Some of the creases become blocked and inactive, i.e., $\rho_{i}=-\pi,0$ or $\pi$, as the folding proceeds. In the simulation, all the creases
are specified \emph{a priori} as shown in Fig.\ref{fig:craneCP}(a).
Fig.\ref{fig:craneCP}(b) shows the paper model for the finial folded crane.
It is worth mentioning that, unlike Randlett's flapping bird \cite{Ghassaei2018FastI}, the current crane is rigid-foldable.

\begin{figure}
	\begin{centering}
		\includegraphics[width=14cm]{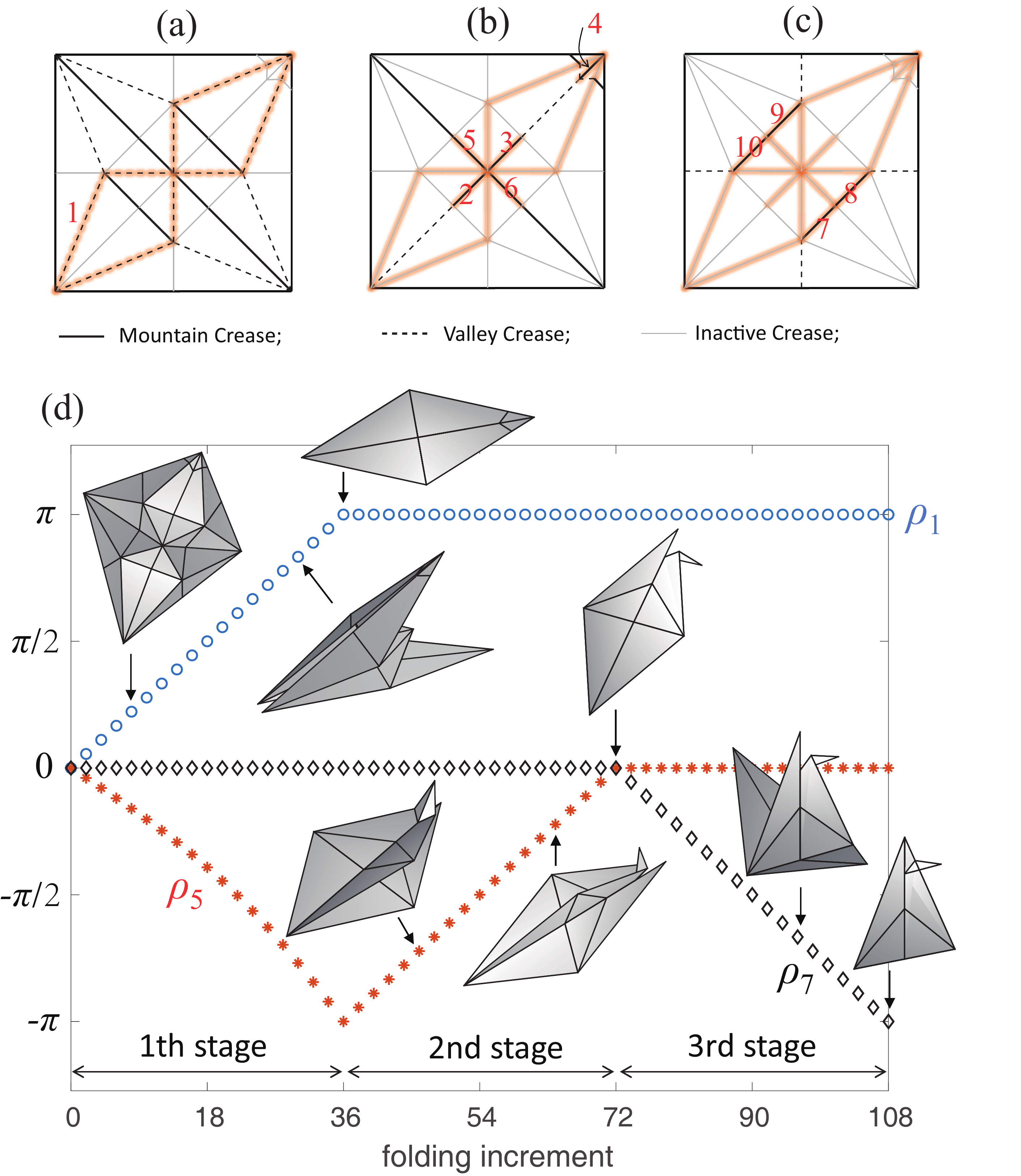}
		\par\end{centering}
	\caption{\label{fig:craneFolding} Sequential folding of origami crane.
		(a), (b) and (c) show the controlled crease patterns for the first, second and third folding substeps, respectively; The foldable creases in solid and dash line are mountains and valleys respectively; the creases under control are highlighted in orange color, and the light gray solid lines are the inactive creases. (d) plots the history of $\rho_1$, $\rho_5$ and $\rho_7$ which correspond to the fold angles at the labeled creases in (a), (b) and (c); the insets show the 3D folded forms during the folding substeps.}
\end{figure}

Through folding by hand, three successive folding stages are identified.
The crease pattern for the first stage is shown in Fig.\ref{fig:craneFolding}(a) where the highlighted creases are controlled.
The gray creases are inactive in this stage, i.e., the pertinent fold angles equal 0 throughout this stage.
Thus, except the central vertex, other vertexes are of degree-4 and are also flat-foldable.
For flat-foldable degree-4 vertex, it is known that the fold angles of opposite creases
are equal in magnitude \cite{evans2015rigidly}.
Besides, the symmetry with respect to the diagonal creases is assumed.
Thus, the fold angles at all the controlled creases are equal and they increase from 0 to $\pi$ by a small amount per folding increment, for instance $5^{\circ}$ in the simulation.
The controlled fold angles $\rho_1$ is chosen as a representative and its history is shown in Fig.\ref{fig:craneFolding}(d).
The fold angles of the uncontrolled creases are calculated by the Algorithm \ref{alg:one_folding}.
At the end of the first folding stage when the controlled fold angles reach $\pi$, the paper is folded flat.
Three frames for the first stage are shown as insets in Fig.\ref{fig:craneFolding}(d).

The crease pattern for the second folding stage is shown in Fig.\ref{fig:craneFolding}(b).
Apart from the controlled creases inherited from the first stage (the highlighted gray creases), the highlighted solid creases numbered from 2 to 6 are also controlled which drive the folding of the second stage.
The fold angles at creases 2, 3 and 4 decrease from 0 to $-\pi$ whilst those at crease 5 and 6 increase from $-\pi$ to 0, see Fig.\ref{fig:craneFolding}(d) for the history of $\rho_5$.
Three frames of the folded form are shown by the insets in Fig.\ref{fig:craneFolding}(d) among which the last one shows that the paper is again folded flat.
As shown in Fig.\ref{fig:craneFolding}(c), despite the controlled fold angles inherited from the first and second stages, four creases numbered from 7 to 10 are added to be controlled which drive the folding of the last stage.
Fig.\ref{fig:craneFolding}(d) shows two frames for the last stage and the history of $\rho_7$.
In the last two stages, it can be seen that more creases become inactive, i.e., more fold angles keep constant at $-\pi,0$ or $\pi$.
To transfer from one stage to the next, the fold angles should be exactly controlled to reach the planar state which is enforced by the Lagrange multipliers on the controlled fold angles.

\subsection{Bistability of Waterbomb base\label{subsec:Waterbomb_base}}

\begin{figure}
	\begin{centering}
		\includegraphics[width=12cm]{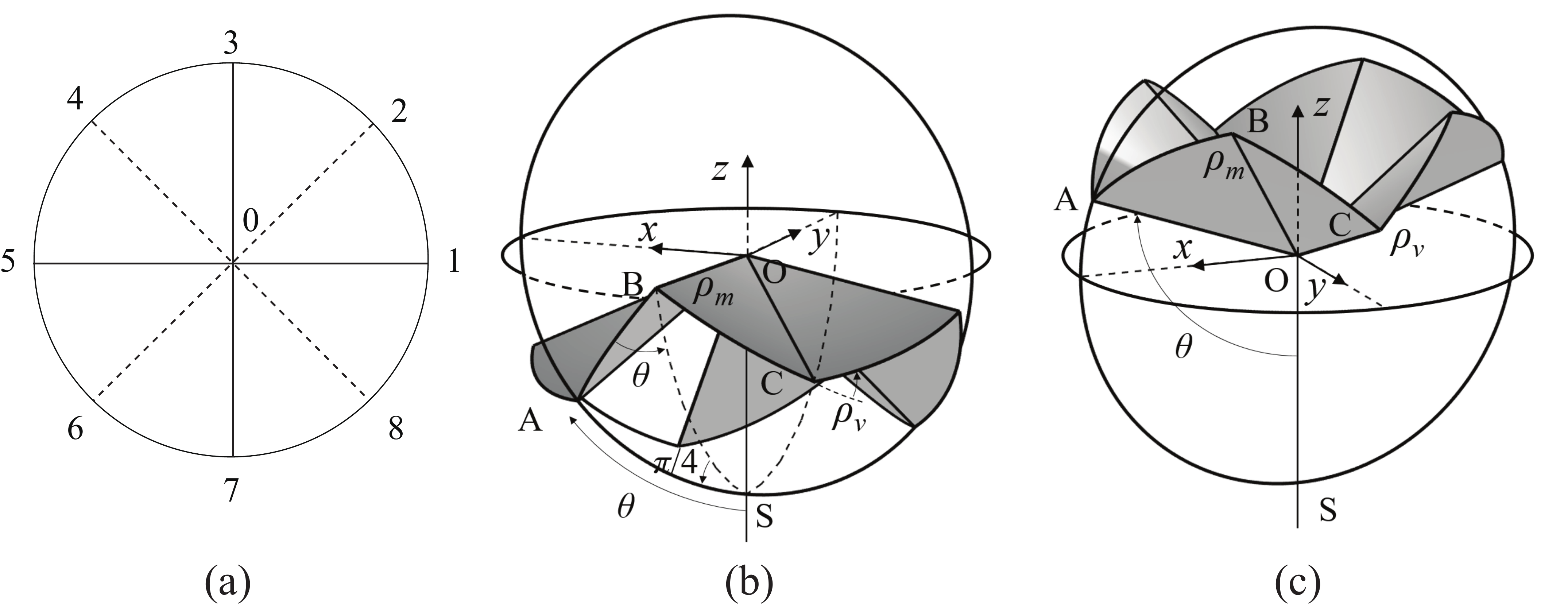}
		\par\end{centering}
	\caption{\label{fig:WB}(a) The crease pattern for the waterbomb base. (b)
		and (c) are the folded form of the bistable states for the case with symmetric rest angles, i.e., all mountain folds are of the same rest angle and so are the valley folds; $\theta$ is the angle between OS and OA.}
\end{figure}

Fig.\ref{fig:WB}(a) shows the circular waterbomb base of unit radius
consisting of eight alternative mountain and valley creases around
a single vertex. It has been suggested as a test bed for actuated
origami systems since the waterbomb base has multiple DOFs and exhibits
bistable behavior \cite{hanna2014waterbomb}. For simplicity, the
spring stiffness per unit length is taken as 1 in the following.

\begin{figure}
	\begin{centering}
		\includegraphics[width=8cm]{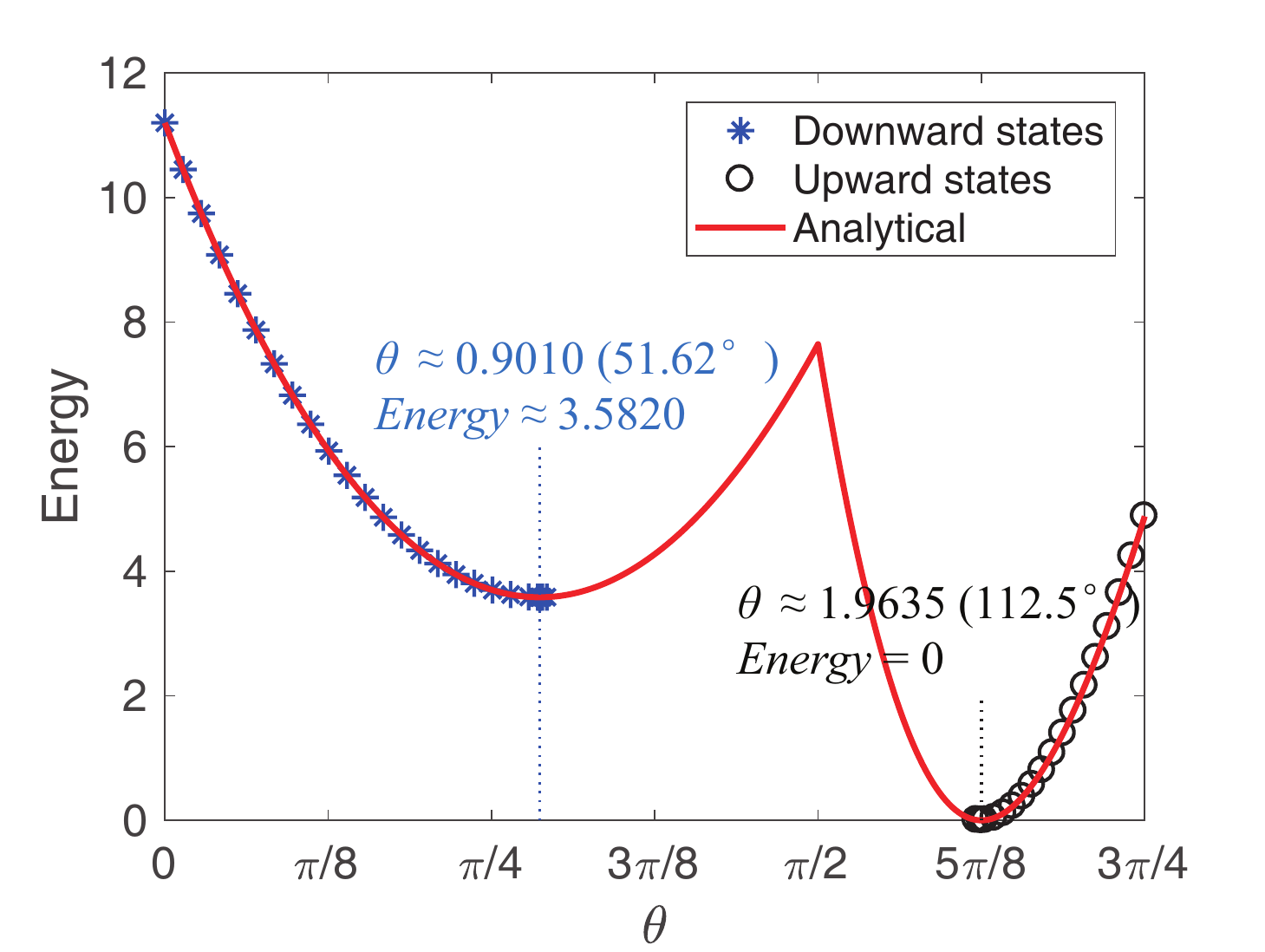}
		\par\end{centering}
	\caption{\label{fig:WB_energy}The energy versus $\theta$ in the searching
		process for the bistable states of the case with symmetric rest angles
		by the Algorithm \ref{alg:one_folding_eng}. The ``{*}'' at $\theta=0$
		(``$\circ$'' at $\theta=3\pi/4$) is for the downward (upward)
		compactly folded state and used as the initial state for the searching
		process.}
\end{figure}

First, we consider the symmetric case in which all mountain folds
are of the same rest angle and so are the valley folds. Thus, the
equilibrated folded forms should also be symmetric. Referring to Fig.\ref{fig:WB}(b),
(c) and using spherical trigonometry, the analytical solutions for
the mountain and valley folds are
\begin{equation}
\left[\rho_{m},\ \rho_{v}\right]=\left\{ \begin{array}{rl}
\left[2\theta-\pi,\ 2\arccos\Big(\dfrac{\sqrt{2}\cos\theta}{-2-\sqrt{2}\sin\theta}\Big)-\pi\right] & 0\leq\theta<\dfrac{\pi}{2}\ ;\\
\left[\pi-2\theta,\ 2\arccos\Big(\dfrac{\sqrt{2}\cos\theta}{2-\sqrt{2}\sin\theta}\Big)-\pi\right] & \dfrac{\pi}{2}\leq\theta\leq\dfrac{3\pi}{4}\ .
\end{array}\right.\label{eq:WBana}
\end{equation}
where $\theta$ is the angle between $\overrightarrow{OS}$ and $\overrightarrow{OA}$.
The analytical solutions, $[\tilde{\rho}_{m},\tilde{\rho}_{v}]=\left[-\pi/4,\ 1.7908\cdots\right]$
obtained by substituting $\theta=5\pi/8$ into Eq.(\ref{eq:WBana}),
are chosen as the rest angles for the mountain and valley creases. The analytical solution for the
energy is readily available from Eq.(\ref{eq:objective_U}). To search the bistable states, the Algorithm \ref{alg:one_folding_eng} is run
twice with two initial states, i.e., the mountain and valley fold
angles are set as $[-\pi,\ \pi/2]$ to search for the stable state
similar to that shown in Fig.\ref{fig:WB}(b) and $[-\pi/2,\ \pi]$
for that shown in Fig.\ref{fig:WB}(c); the two initial states correspond
to the downward compactly folded state with $\theta=0$ and the upward
compactly folded state with $\theta=3\pi/4$, respectively. The searching
paths of the Algorithm \ref{alg:one_folding_eng} are indicated by
the ``{*}''s and ``$\circ$''s in Fig.\ref{fig:WB_energy}. The
folded forms of the bistable states are similar to those in Fig.\ref{fig:WB}(b)
and (c) and thus are not plotted. The predictions for the fold angles
and energy of the equilibrium states are in agreement with the analytical
solution.

\begin{figure}
	\begin{centering}
		\includegraphics[width=13cm]{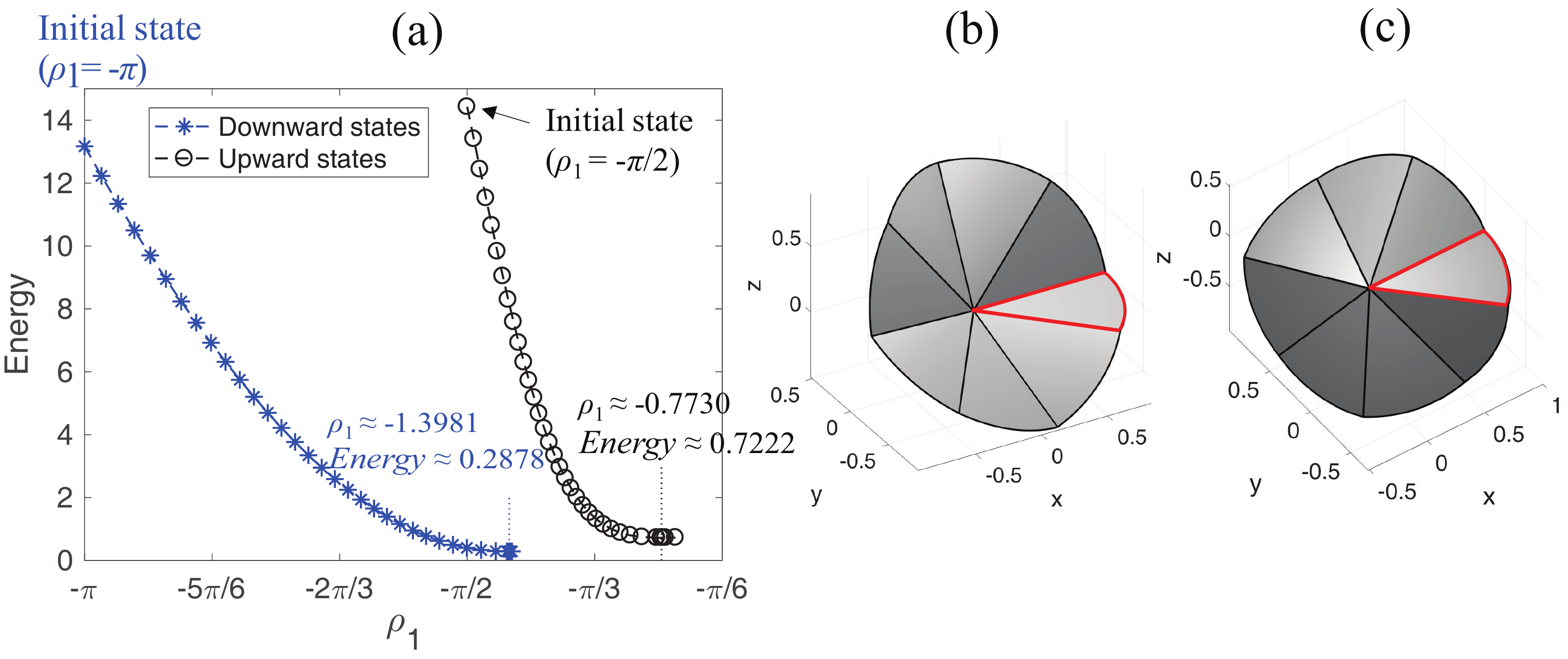}
		\par\end{centering}
	\caption{\label{fig:WB_bi}The waterbomb base with the unsymmetric rest angles:
		(a) the searching paths by Algorithm \ref{alg:one_folding_eng} for
		the downward and upward stable states are indicated by ``{*}''s
		and ``$\circ$''s, respectively; (b) and (c) are respectively the
		folded forms for the downward and upward stable states where the facet
		0-1-2 with the red edges are kept stationary.}
\end{figure}

Next, we chose the rest angles as $\tilde{\rho}_{i}=(-1)^{i}\frac{\pi}{i+1}$
for crease-$i$ with $i=1,2,\cdots,8$. Analytical solutions are not available for this case.
Similar to the symmetirc case, the two
initial states, i.e., downward and upward compactly folded states,
are used to search the bistable states. As the states in the searching
process are not symmetric, the angle $\theta$ is not well defined.
Instead, the searching path of the Algorithm \ref{alg:one_folding_eng}
is indicated by the energy against the characteristic fold angle $\rho_{1}$
shown in Fig.\ref{fig:WB_bi}(a). The folded forms of the bistable states
are shown in Fig.\ref{fig:WB_bi}(b) and (c), respectively.

\subsection{Elastic equilibrium configuration of Waterbomb tessellation\label{subsec:WBT}}

\begin{figure}
	\begin{centering}
		\includegraphics[width=7cm]{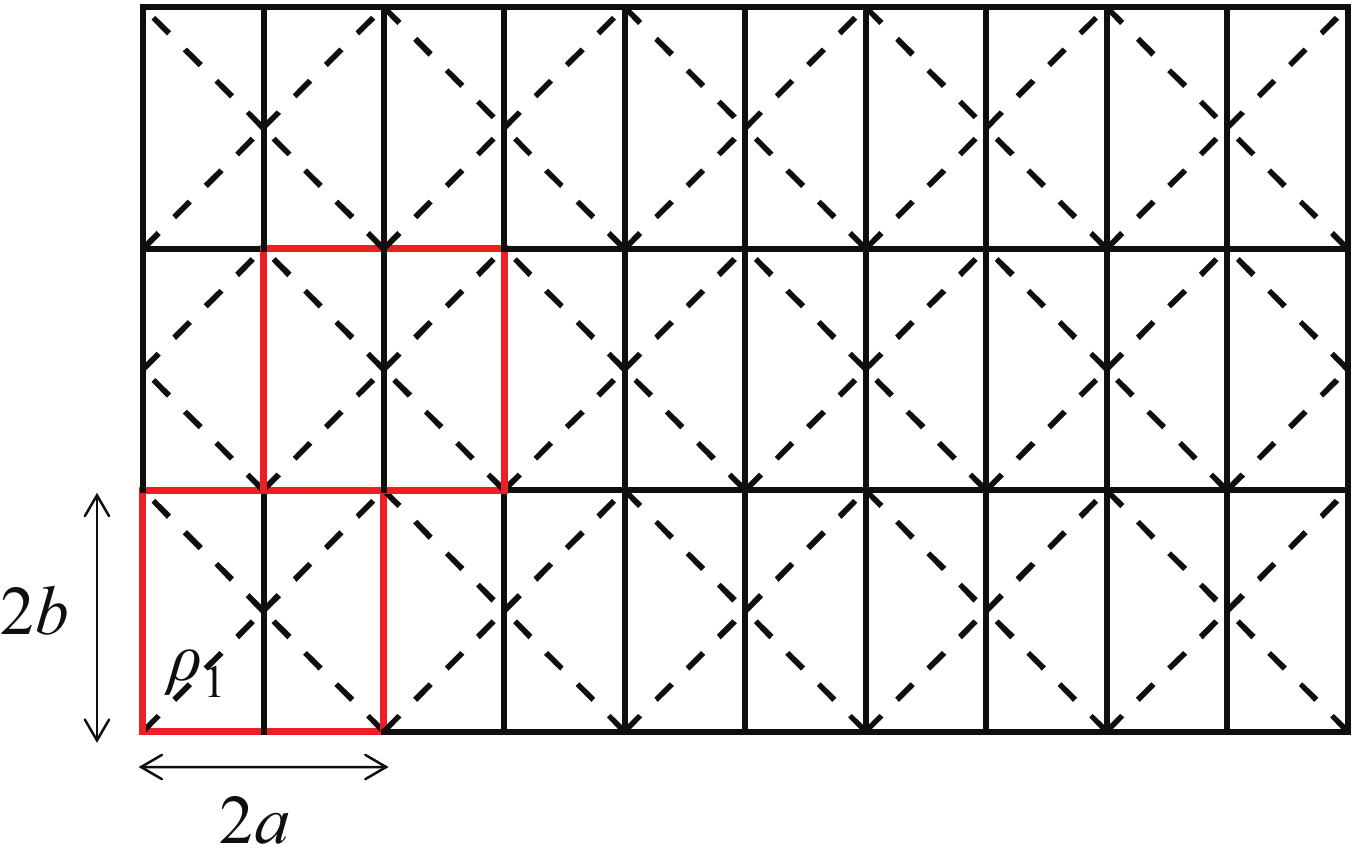}
		\par\end{centering}
	\caption{\label{fig:SWBcp} The crease pattern for the Waterbomb tessellation
		made up of 5$\times$3 bases with the lengths $a=b=1$. }
\end{figure}

\begin{figure}
	\begin{centering}
		\includegraphics[width=13cm]{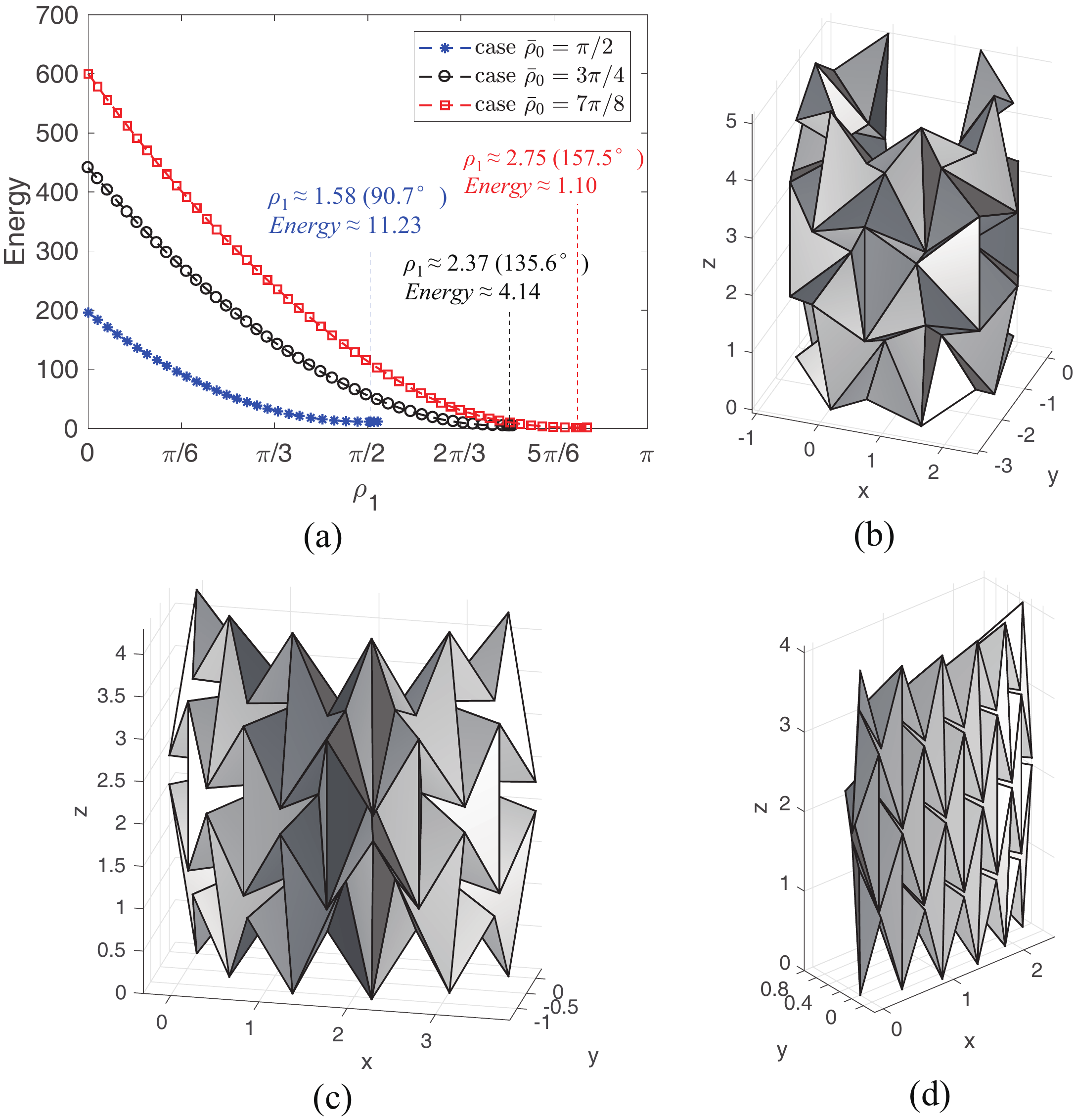}
		\par\end{centering}
	\caption{\label{fig:SWB3case}The searching paths (a) and the equilibrium configurations
		(b), (c) and (d) for the cases with $\tilde{\rho}_{0}=\pi/2,3\pi/4$
		and $7\pi/8$ by the Algorithm \ref{alg:one_folding_eng}. }
\end{figure}

\begin{figure}
	\begin{centering}
		\includegraphics[width=13cm]{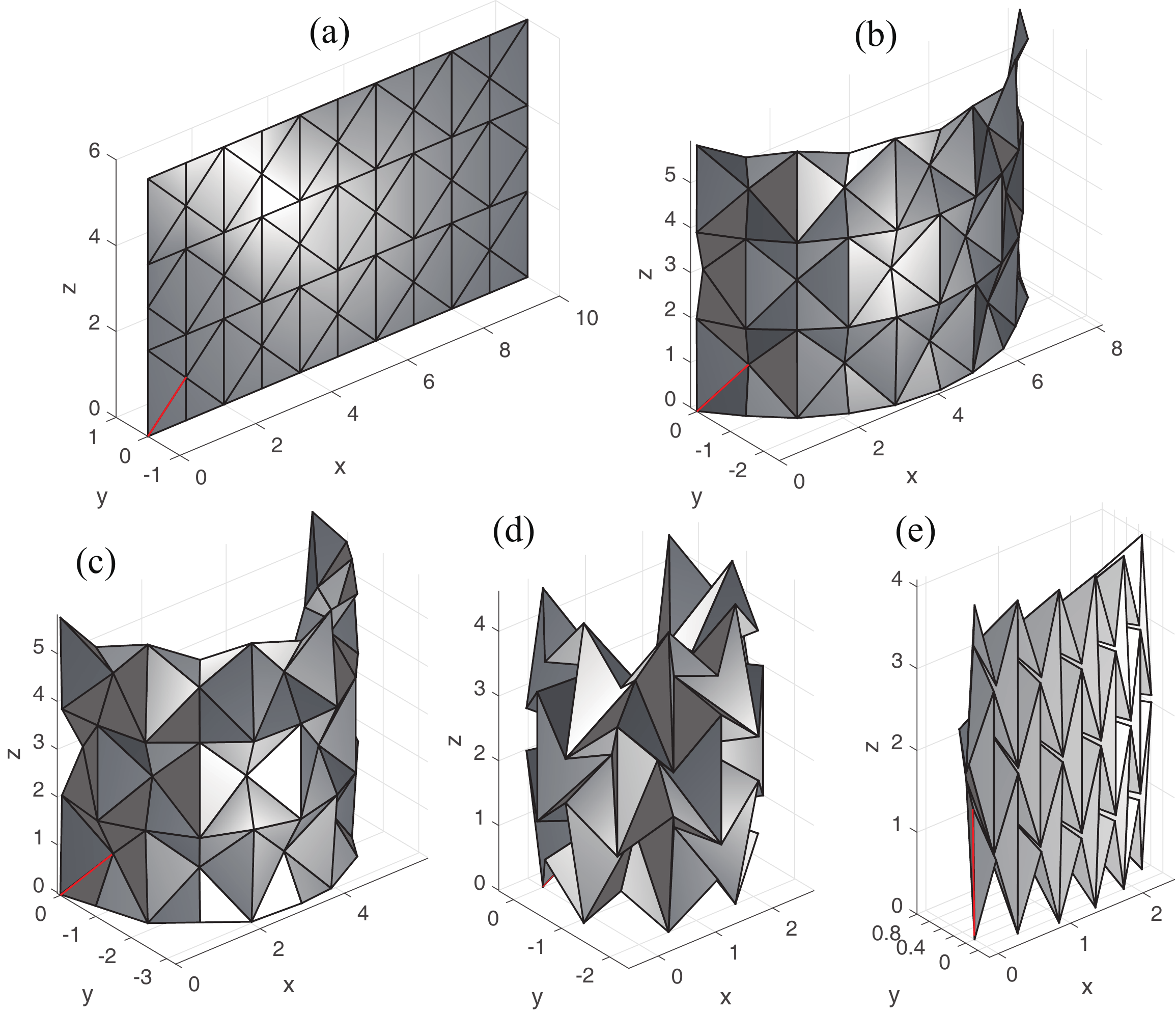}
		\par\end{centering}
	\caption{\label{fig:SWBfolding}Snapshots of the intermediate states for the
		case with $\tilde{\rho}_{0}=7\pi/8$ by the Algorithm \ref{alg:one_folding_eng};
		(a) through (e) correspond respectively to the states with $\rho_{1}=0^{\circ},28.5^{\circ},44.6^{\circ},113.6^{\circ}$
		and $157.5^{\circ}$ where $\rho_{1}$ is indicated by the red line. }
\end{figure}

This example considers the waterbomb tessellation made up of degree-6
bases, see the square with red edges in Fig.\ref{fig:SWBcp}. The
rotational stiffness per unit length is taken as $k=1$. The rest
angles for the mountain and valley creases are, respectively, set
as $\tilde{\rho}_{v}=-\tilde{\rho}_{0}$ and $\tilde{\rho}_{m}=\tilde{\rho}_{0}$
and three cases with $\tilde{\rho}_{0}=\pi/2,3\pi/4$ and $7\pi/8$
are studied. The initial fold angles are 0 and the fold angle $\rho_{1}$
indicated in Fig.\ref{fig:SWBcp} is used to characterize the folded
states during the searching process. Fig.\ref{fig:SWB3case}(a) shows
the searching paths of Algorithm \ref{alg:one_folding_eng} for the
three cases and the corresponding equilibrium configurations are in
Fig.\ref{fig:SWB3case}(b), (c) and (d). It can be seen that the waterbomb
tessellation curves into a tube for $\tilde{\rho}_{0}=\pi/2$ and
flattens out as $\tilde{\rho}_{0}$ increases. The computational costs
of the three cases are, respectively, 1.8s, 2.3s and 2.5s for the
Algorithm \ref{alg:one_folding_eng}.\footnote{For reference, the MATLAB implementation was run on a desktop with
Intel(R) Core(TM) i7-6700 (8 cores, 3.41Gz).} As the residual of each step is negligible, the intermediate states
of the Algorithm \ref{alg:one_folding_eng} are valid and can be visualized;
the snapshots for the states with $\rho_{1}=0^{\circ},28.5^{\circ},44.6^{\circ},113.6^{\circ}$
and $157.5^{\circ}$ are shown in Fig.\ref{fig:SWBfolding}(a) through
(e), respectively.

\section{Closure}

In this paper, we propose algorithms for rigid origami folding analysis using fold angles as variables.
By combining the loop closure constraint with Lagrange multiplier method, we are able to model sequential origami folding when the crease pattern is of multiple DOFs.
The introduction of Lagrange multiplier method allows us to control the fold angles of different sets of creases, so that some creases fold and the rest remain fixed at successive substeps.
Newton-Raphson method is adopted in the algorithm to eliminate the numerical residual, and the geometric features of the origami can be accurately traced during the sequential folding simulations.
This strategy is also extended to model rigid origami with rotational springs mounted at the creases, which involves elastic energy cost and competition between creases during the folding process.
To find the equilibrium configurations of origami with elastic rotational springs of different rest angles, we construct a functional to minimize the elastic spring energy while enforcing the loop closure constraint with Lagrange multiplier method.

The two algorithms are applicable to general origami structures without holes, particularly useful for the rigid origami with irregular unit cells and multiple DOFs in which cases the analytical geometric analysis can be tedious.
Many origami artworks involve sequential folding, and they are continulously inspiring the development of packaging, reconfigurable electronics, self-folding robotics, etc.
Those origaim-inspired applications often harness sequential folding \cite{lee2015sequential,liu2017sequential} or elastic folding drivend by actuators embedded at the creases \cite{hawkes2010programmable,ge2014active,rus2018design}, where the present algorithms are applicable for the kinematic simulations.

Since the algorithms are based on rigid origami, they are not suitable for simulating elastically deformed origami, when facets are generally under stretching, shearing and bending. 
Another limitation of the algorithms is that the sequential folding algorithm relies on prior known folding sequence. It is attractive to search the folding sequence for a given crease pattern, which will be explored in our future works.

\appendix

\section{Loop closure constraint\label{sec:append_A}}

This appendix first derives the loop closure constraint; then the
derivative of the constraint matrix with respect to fold angles is
proved to be antisymmetric for compatible fold angles. In Fig. \ref{fig:vertexfan}(b),
the triplet, $\textbf{L}_{i}$, $\textbf{M}_{i}$ and $\textbf{N}_{i}$,
defines a local Cartesian coordinate system for the sector facet $i$-$(i$+1)
with $\textbf{L}_{i}$ being the unit vector along crease $i$, $\textbf{N}_{i}$
being the unit normal of the sector facet $i$-$(i+1)$ and $\textbf{M}_{i}=\textbf{N}_{i}\times\textbf{L}_{i}$
where ``$\times$'' is the cross product. It is clear that
\begin{equation}
[\textbf{L}_{i},\,\textbf{M}_{i},\,\textbf{N}_{i}]=[\textbf{L}_{i-1},\,\textbf{M}_{i-1},\,\textbf{N}_{i-1}]\bm{\chi}_{i-1,i}\label{eq:coodRot}
\end{equation}
where $\bm{\chi}_{i-1,i}$ given in Eq.(\ref{eq:single_rot}) for
$i=1,2,\cdots,n$ and ($i$-1) is to be interpreted cyclically. Starting
from the sector facet 1-2 and looping around the vertex anticlockwise,
recursive usage of the transformation in Eq.(\ref{eq:coodRot}) yields
\begin{equation}
[\textbf{L}_{1},\textbf{M}_{1},\textbf{N}_{1}]=[\textbf{L}_{1},\textbf{M}_{1},\textbf{N}_{1}]\bm{\chi}_{1,2}\bm{\chi}_{2,3}\cdots\bm{\chi}_{n,1}\ .
\end{equation}
Thus, the loop closure constraint is obtained as

\begin{equation}
\textbf{F}(\bm{\rho}_{P})=\bm{\chi}_{1,2}\bm{\chi}_{2,3}\cdots\bm{\chi}_{n,1}=\textbf{I}_{3}\ .\label{eq:LoopCons_ap}
\end{equation}

Since $\chi_{i,j}$'s are orthogonal, $\textbf{F}\textbf{F}^{\text{T}}$
always yield the identity matrix and its derivative with respect to
any fold angle vanish. With Eq.(\ref{eq:LoopCons_ap}), it is clear
that \cite{tachi2009simulation}

\[
\left.\frac{\partial(\textbf{F}\textbf{F}^{\text{T}})}{\partial\rho_{i}}\right|_{\bm{\rho}_{P}}=\left.\left[\frac{\partial\textbf{F}}{\partial\rho_{i}}\textbf{F}^{\text{T}}+\textbf{F}\frac{\partial(\textbf{F}^{\text{T}})}{\partial\rho_{i}}\right]\right|_{\bm{\rho}_{P}}=\left.\frac{\partial\textbf{F}}{\partial\rho_{i}}\right|_{\bm{\rho}_{P}}+\left.(\frac{\partial\textbf{F}}{\partial\rho_{i}})^{\text{T}}\right|_{\bm{\rho}_{P}}=\textbf{0}\ .
\]
In other words, the matrix $\frac{\partial\textbf{F}}{\partial\rho_{i}}$
is antisymmetric when the fold angles are compatible. Besides, from
Eq.(\ref{eq:LoopCons_ap}) and (\ref{eq:single_rot}), we have
\begin{equation}
\frac{\partial\textbf{F}}{\partial\rho_{i}}=\bm{\chi}_{1,2}\cdots\bm{\chi}_{i-2,i-1}\frac{\partial\bm{\chi}_{i-1,i}}{\partial\rho_{i}}\bm{\chi}_{i,i+1}\cdots\bm{\chi}_{n,1}\label{eq:analy_dF}
\end{equation}
where
\[
\frac{\partial\bm{\chi}_{i-1,i}}{\partial\rho_{i}}=\begin{bmatrix}\cos\theta_{i-1} & -\sin\theta_{i-1} & 0\\
\sin\theta_{i-1} & \cos\theta_{i-1} & 0\\
0 & 0 & 1
\end{bmatrix}\begin{bmatrix}1 & 0 & 0\\
0 & -\sin\rho_{i} & -\cos\rho_{i}\\
0 & \cos\rho_{i} & -\sin\rho_{i}
\end{bmatrix}.
\]
Thus, the entries of $\frac{\partial\textbf{F}}{\partial\rho_{i}}$,
i.e., $a_{i},b_{i}$ and $c_{i}$ in Eq.(\ref{eq:asymF}), can be
obtained analytically from Eq.(\ref{eq:analy_dF}).

\section{Folded form for given crease pattern and fold angles\label{sec:append_B}}

\begin{figure}
	\begin{centering}
		\includegraphics[width=7cm]{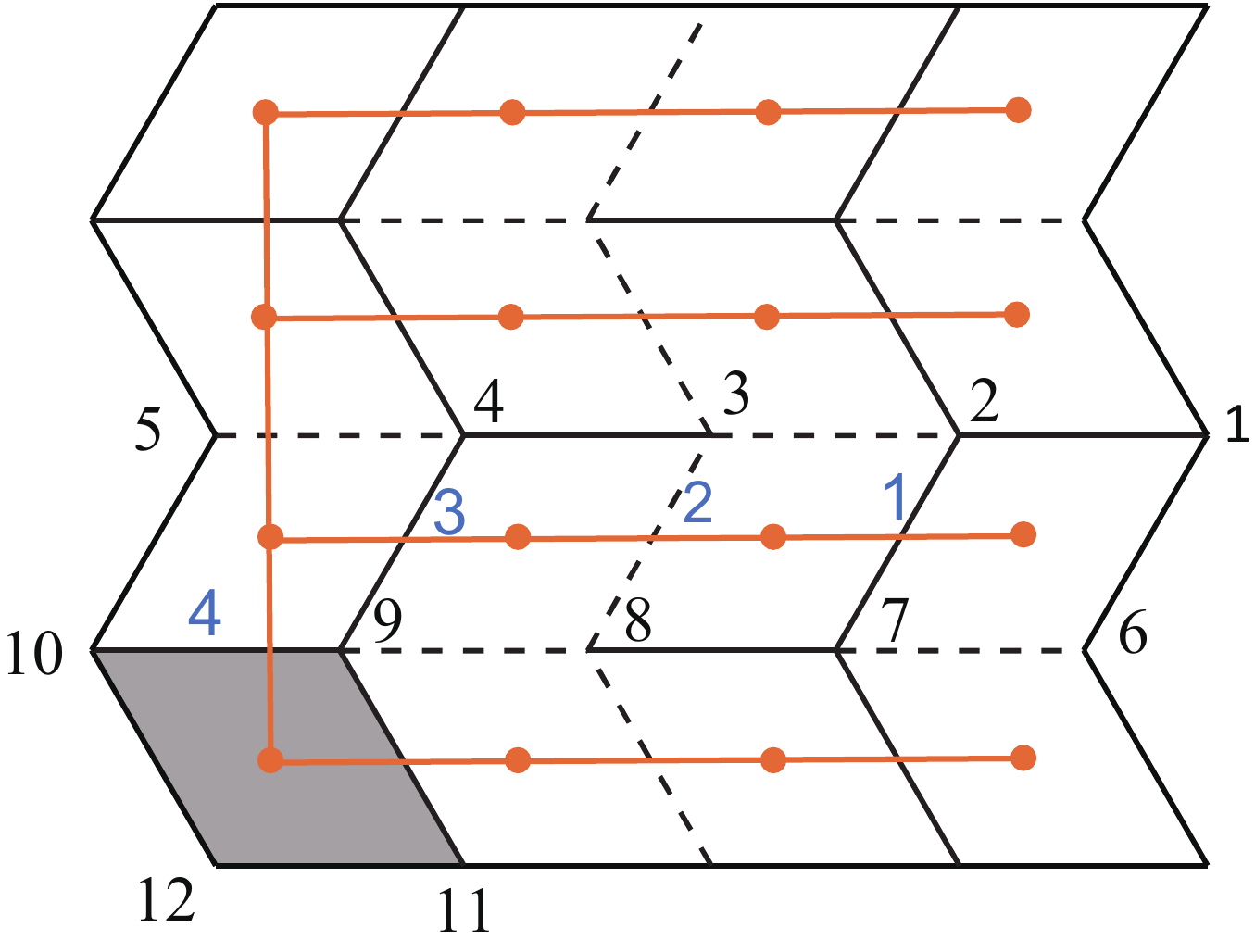}
		\par\end{centering}
	\caption{\label{fig:appendfig}The spanning tree for the calculation of the
		vertex coordinates: the gray facet is the root and the paths for calculating
		the vertexes coordinates through successive rotation are indicated
		by the yellow lines.}
\end{figure}

When the crease pattern and fold angles are known, the shape of folded
form is determined. However, the position and orientation of the folded
form are undetermined which can be fixed by specifying the position
of one of the facets. A natural choice is to set the first facet stationary,
i.e., the vertex $x$- and $y$-coordinates of the first facet are
the same as those in the crease pattern while the $z$-coordinates
are set as 0. The other facets are calculated through successive rotations
with respect to the pertinent creases by the fold angles. To ensure
an ordered calculation, a spanning tree can be constructed on the
crease pattern with its root at the facet whose position is specified
\cite{lang2017twists}, see Fig.\ref{fig:appendfig} for an illustration.
For instance, we consider the calculation of the coordinates of vertex-1
whilst only the relevant vertexes (black numbers) and creases (blue
numbers) are indicated in Fig.\ref{fig:appendfig}. First, the facet
7-6-1-2 is to be rotated with respect to facet 8-7-2-3. The axis of
rotation is the vector $\overrightarrow{V_{2}V_{7}}$. Due to the
sign conventions of the fold angles in Fig.\ref{fig:vertexfan}(a),
the order of the vertexes for the axis should be in the clockwise
direction regarding the facet 8-7-2-3 which is nearer to the root
facet 12-11-9-10. Let $\rho_{i}$ be the fold angle at crease-$i$;
$\{ai,bi\}$ be the pair of ordered vertex numbers for crease-$i$
such that $\overrightarrow{V_{ai}V_{bi}}$ is in the clockwise direction
of the facet that is near the stationary facet; and $\textbf{X}_{i}=\{x_{i},y_{i},0\}$
be the coordinates of the crease pattern vertex-$i$, then
\begin{equation}
\textbf{x}_{1}=\textbf{R}_{a}(\rho_{i},\textbf{e}_{i})\big(\textbf{x}_{1}-\textbf{X}_{ai}\big)+\textbf{X}_{ai}\quad\text{with}\quad\textbf{e}_{i}=\frac{\textbf{X}_{bi}-\textbf{X}_{ai}}{|\textbf{X}_{bi}-\textbf{X}_{ai}|}\label{eq:rotV}
\end{equation}
where $\textbf{R}_{a}(\rho_{i},\textbf{e}_{i})$ is the rotation matrix
which rotate a vector by angle $\rho_{i}$ about the unit vector $\textbf{e}_{i}$,
see \cite{lang2017twists} for the expressions and $\textbf{x}_{i}=\{x,y,z\}$
donates the coordinates of vertex-$i$ in the 3D space. The rotation
can be repeated until the stationary facet 12-11-9-10. When all the
nodal coordinates are obtained, the 3D folded form can be rotated
and translated to meet certain conditions or just for view effect.

\bibliography{mybibfile}

\end{document}